\def\Black{} 
\def\Blue{} 
\def\Brown{} 
\documentstyle[prl,aps,epsfig]{revtex} 
\begin{document} 
 
\def \gam {\frac{ N_f N_cg^2_{\pi q\bar q}}{8\pi} } 
\def \gamm {N_f N_cg^2_{\pi q\bar q}/(8\pi) } 
\def \be {\begin{equation}} 
\def \ba {\begin{eqnarray}} 
\def \ee {\end{equation}} 
\def \ea {\end{eqnarray}} 
\def \gap {{\rm gap}} 
\def \gapp {{\rm \overline{gap}}} 
\def \gappp {{\rm \overline{\overline{gap}}}} 
\def \im {{\rm Im}} 
\def \re {{\rm Re}} 
\def \Tr {{\rm Tr}} 
\def \P {$0^{-+}$} 
\def \S {$0^{++}$} 
\def \uu {$u\bar u$} 
\def \dd {$d\bar d$} 
\def \ss {$s\bar s$} 
\def \qq {$q\bar q$} 
\def \qqq {$qqq$} 
\def \si {$\sigma(500-600)$} 
\def \lsm {L$\sigma $M} 
\begin{titlepage} 
\null 
\noindent 
\flushright{CERN-TH/2000-242}\\ 
\vspace{2.5cm} 
\begin{center} 
\Brown 
\Large\bf 
$B\to\rho\pi$ decays, resonant and nonresonant contributions 
\Black 
\end{center} 
\vspace{1.5cm} 
 
\begin{center} 
\begin{large} 
A. Deandrea$^a$ and A.D.~Polosa$^{b}$\\ 
\end{large} 
\vspace{0.7cm} 
$^a$ Theory Division, CERN, CH-1211 Gen\`eve 23, Switzerland\\ 
$^b$  Physics Department, POB 9, FIN--00014, University of Helsinki, 
Finland\\ 
\end{center} 
 
\vspace{2cm} 
 
\begin{center} 
\begin{large} 
\Brown 
{\bf Abstract}\\[0.5cm]\Black 
\end{large} 
\parbox{14cm}{We point out that a new contribution to B decays to three pions 
is relevant in explaining recent data from 
the CLEO and BABAR collaborations, in particular the results 
on quasi--two--body decays via a $\rho$ meson. We also discuss the 
relevance of these contribution to the measurement of CP 
violations.} 
\end{center} 
 
\vspace{3cm} 
\noindent 
\Blue 
\Black 
 
\noindent 
August 2000 
\end{titlepage} 
 
\setcounter{page}{1} 
 
\twocolumn 
 
\title{$B\to\rho\pi$ decays, resonant and nonresonant contributions} 
\author{A. Deandrea$^{(a)}$ and A.D. Polosa$^{(b)}$} 
\address{(a) Theory Division, CERN, CH-1211 Gen\`eve 23, 
Switzerland\\ 
(b) Physics Department, POB 9, FIN--00014, University of Helsinki, 
Finland} 
\date{August, 2000} 
\maketitle 
 
\begin{abstract} 
We point out that a new contribution to B decays to three pions 
is relevant in explaining recent data from 
the CLEO and BABAR collaborations, in particular the results 
on quasi--two--body decays via a $\rho$ meson. We also discuss the 
relevance of these contribution to the measurement of CP 
violations. 
\end{abstract} 
 
\preprint{CERN-TH/2000-242} 
 
Several exclusive charmless hadronic $B$ decays are known 
with good accuracy, and more will be measured in the near future, due 
to the large amount of data coming from $e^+e^-$ machines such as 
the CLEO experiment \cite{cleo} at the Cornell Electron Storage 
Ring (CESR), the BaBar experiment at SLAC \cite{babar} and the BELLE 
experiment at KEK \cite{kek}, or hadron machines such as the 
Large Hadron Collider (LHC) at CERN, with its program for 
$B$--physics. In the following we will consider quasi--two--body 
$B$ decays to three pions. Our motivation is twofold. 
 
(1) ${\bar B}^0 \to \rho^\mp \pi^\pm$ and $B^- \to \rho^0 \pi^-$ 
were recently measured by the CLEO  and BABAR 
collaborations. The ratio of these branchings 
are consistent among the two experiments and different from 
the theoretical expectations. 
 
(2) quasi--two--body $B$ decays to three $\pi$ can be used 
for the determination of the angles of the unitarity triangle 
of CP violations \cite{cp}. 
 
Concerning the first point, the CLEO collaboration finds 
\cite{cleob} $ {\mathcal B}( 
B^-~\to~\rho^0\pi^-)~=~(10.4^{+3.3}_{-3.4}\pm 2.1)\times 10^{-6}$ 
and ${\mathcal B}( {\bar 
B}^0~\to~\rho^\mp\pi^\pm)~=~(27.6^{+8.4}_{-7.4}\pm 4.2)\times 
10^{-6}$, combining these numbers we find a ratio 
\begin{equation} 
\label{ratio} 
R~=~\frac{ {\mathcal B}( {\bar B}^0~\to~\rho^\mp\pi^\pm) }{ {\mathcal B}( 
B^-~\to~\rho^0\pi^-)}~=~2.65\pm 1.8 \; , 
\end{equation} 
where the error in the ratio includes all errors from the branchings; the 
error may be smaller than what indicated if part of the systematics 
simplifies in the ratio. The BABAR collaboration \cite{babarb} gives the 
following preliminary results 
$ {\mathcal B}( B^-~\to~\rho^0\pi^-)~=~(24\pm 8\pm 
3)\times 10^{-6}$ and ${\mathcal B}( {\bar 
B}^0~\to~\rho^\mp\pi^\pm)~=~(49\pm13^{+6}_{-5})\times 
10^{-6}$, with these numbers we find a ratio $R=2.0 \pm 1.3$. 
 
As discussed in \cite{gao}, this ratio is rather small with respect to 
theoretical expectations; as a matter of fact, when computed in simple 
approximation schemes, as factorization with no penguins, one gets, 
from the BSW model \cite{BSW}, $R \simeq 6$. 
A calculation including penguins in the factorization 
approximation was performed in \cite{bstar1} 
and gives $R \simeq 5.5$. Calculations beyond factorization were performed 
in \cite{martinelli} with similar results \cite{bstar2} (see Table I). 
 
Previous papers \cite{bstar1,bstar2,bstar3,desh} investigated to role of the 
$B^*$ and higher resonances in these decays, while here we will investigate 
a completely different mechanism that enhances selectively some 
of the $B \to 3 \pi$ decays, namely the possibility that a 
broad light scalar resonance is present in the 3--body Dalitz 
plot. 
 
The light $\sigma$ resonance has accumulated considerable 
interest after it was reintroduced as a very broad resonance into 
the 1996 edition of the Reviews of Particle Physics \cite{pdg}. 
The E791 collaboration has an evidence of a very broad scalar resonance 
\cite{E791} having mass $m_\sigma=478\pm 24$ MeV and width 
$\Gamma_\sigma=324\pm 41$ MeV taking part into the decay process 
$D^+\to 3\pi$ via the resonant channel 
$D^+\to\sigma\pi^+\to 3\pi$. The broad $\sigma$ was 
controversial due to the extreme difficulty in disentangling 
it from available data on $\pi\pi\to\pi\pi$. In the following 
we will use the the experimental numbers for the mass and width of 
the $\sigma$ given by the E791 collaboration.  
In the process $D^+\to 3\pi$ the $\sigma$ is a good description of 
three $\pi$ events not due to quasi--two--body decays of a narrow 
resonance as it accounts for $46\%$ of the three $\pi$'s branching 
\cite{analisi}. 
We will model $B^+\to 3\pi$ in a similar way and see that indeed such a 
contribution is relevant and larger than or comparable to the non--resonant 
long--distance contribution of the $B^*$. These 
two contributions, after the appropriate experimental 
cuts, constitute an irreducible background to the processes 
$B \to \rho \pi$, therefore adding events to some of them and 
modifying the ratio of branching ratios. This has also consequences 
on methods based on $B \to \rho \pi$ for the determination of the 
CP violating angles. It should be mentioned that 
the latest edition of the Reviews of Particle Physics \cite{pdg} 
gives a broad range for the $\sigma$ mass and width. Their precise 
value is not crucial to our analysis as long as the 
the sigma is within the strip of the Dalitz plot where 
the $\rho$ mass contribution (within few hundred MeV due to the 
experimental cuts) is found. 
 
The amplitude we are 
interested in are those corresponding to the diagrams in Figs. 1 
and 2. We compute them at the tree level using the following 
effective Hamiltonian \cite{BSW}: 
\begin{eqnarray} 
H_{\rm eff}&=&\frac{G_F}{\sqrt{2}}V^*_{ub}V_{ud}\{ a_1 
(\bar{u}b)_{V-A}(\bar{d}u)_{V-A}\nonumber \\ 
&+& a_2 (\bar{d}b)_{V-A}(\bar{u}u)_{V-A} \}, \label{eq:uno} 
\end{eqnarray} 
giving, in the first case: 
\begin{eqnarray} 
&\langle& \sigma \pi^-|H_{\rm eff}|B^-\rangle = 
\frac{G_F}{\sqrt{2}}V^*_{ub}V_{ud}a_1 
F_0^{(B\sigma)} m_\pi^2 (m_B^2-m_\sigma^2) \nonumber\\ 
&\times & 
\frac{f_\pi  g_{\sigma \pi^+\pi^-}}{\sqrt{2}} 
\left(\frac{1}{u-m_\sigma^2+i\Gamma_\sigma(u)m_\sigma}+\frac{1} 
{t-m_\sigma^2+i\Gamma_\sigma(t)m_\sigma}\right) \nonumber \\ 
&& 
\label{ff1} 
\end{eqnarray} 
where the calculation of the form factor $F_0^{(B\sigma)}$ 
proceeds as in \cite{pred}, $g_{\sigma\pi^+\pi^-}=2.52$ GeV comes 
from the experimental value of $\Gamma_\sigma$ and the co-moving 
width $\Gamma_\sigma(x)$ is 
\begin{equation} 
\Gamma_\sigma(x)=\Gamma_\sigma 
\frac{m_\sigma}{\sqrt{x}}\frac{\sqrt{x/4-m_\pi^2}} 
{\sqrt{m_\sigma^2/4-m_\pi^2}}, 
\end{equation} 
where $x$ is $u=(p-p_1)^2$ or $t=(p-p_2)^2$ (as in 
the crossed channel, see Fig. 1, in which we have two identical 
$\pi^-$ particles in the final state). The comoving width can be 
obtained comparing the usual formula for the fixed width (see 
for example \cite{pdg}) with a similar formula where the $\sigma$ mass 
is replaced by the square root of the relevant Mandelstam variable 
(as the $\sigma$ is an intermediate state in the decay process). 
The coefficient $a_1$ is 
given by $a_1=C_1+\frac{1}{3}C_2$ where the Wilson coefficients 
$C_1$ and $C_2$, fitted for $B$ decays, are $C_1(m_b)=1.105$ and 
$C_2(m_b)=-0.228$. 
 
On the other hand, the diagram in Fig. 2 is controlled by the 
$a_2$ coefficient appearing in (\ref{eq:uno}) and the result is: 
\begin{eqnarray} 
\langle \sigma \pi^0|H_{\rm eff}|\bar{B}^0\rangle& =& 
\frac{G_F}{\sqrt{2}}V^*_{ub}V_{ud}a_2 
F_0^{(B\sigma)}(m_\pi^2)(m_B^2-m_\sigma^2)f_\pi \nonumber \\ 
&\times& \frac{g_{\sigma\pi^+\pi^-}} 
{u-m_\sigma^2+i\Gamma_\sigma(u)m_\sigma}\; , 
\label{ff2} 
\end{eqnarray} 
with $a_2=C_2+\frac{1}{3}C_1$. We have checked that the formulas 
(\ref{ff1}) and (\ref{ff2}), when applied to $D \to \sigma \pi$ 
with the obvious changes in the masses, form factor values and 
CKM coefficients, reproduce the experimental results of \cite{E791}. 
 
The definition of the form factors $F_0^{(B\sigma)}$ and 
$F_1^{(B\sigma)}$ is the following: 
\begin{eqnarray} 
\langle \sigma(q_\sigma)&|&A^{\mu}(q)|B(p)\rangle = 
\left[\frac{m_D^2-m_{\sigma}^2}{q^2}q^{\mu} \right] \; 
F_0^{(B\sigma)}(q^2)\nonumber \\ 
&+&\left[ 
(p+q_\sigma)^{\mu}-\frac{m_D^2-m_{\sigma}^2}{q^2}q^\mu \right]\; 
F_1^{(B\sigma)}(q^2), 
\label{eq:effezero} 
\end{eqnarray} 
with $F_1(0)=F_0(0)$. Applying the factorization selects the form 
factor $F_0^{(B\sigma)}$ since: 
\begin{equation} 
\langle \pi^-|A^\mu_{(\bar{d}u)}|{\rm VAC}\rangle=if_\pi q^\mu. 
\end{equation} 
The {\it polar} and {\it direct} contributions to the semileptonic 
form factors discussed in \cite{pred} for the $D\to \sigma$ 
transition, give in this case: 
\begin{equation} 
F_0^{(B\sigma)}(m_\pi^2)\simeq F_0(0)= 0.45 \pm 0.15, 
\end{equation} 
where $m_B=5.27$ GeV, $m_B(1^+)=5.7$ GeV, $B(1^+)$ being the polar 
intermediate state connecting $B$ to $A^\mu$ \cite{pred}. The 
errors include the uncertainty due to the variation of the 
parameters of the CQM model \cite{cqm} in a fixed range of values 
and that arising from the extrapolation of the polar form factors 
to $q^2=m_\pi^2\simeq 0$, following the same steps as in 
\cite{pred}. In that paper the analogous form factor for $D$ 
decays was evaluated, $F_0^{(D\sigma)}$, and the result of the 
model closely reproduces the experimental value of \cite{E791}. 
The calculation does not include $1/m_b$ corrections, where $m_b$ 
is the heavy quark mass, however such corrections were estimated 
to be smaller than the quoted experimental error for $D$ decays 
and are even smaller in the study of $B$ decays. 
 
The possibility of a $\sigma$ contribution to CP violations was  
studied in $K \to \pi \pi$ decays \cite{sanda}. We consider in the  
following the impact of $\sigma$ contribution in quasi two--body 
pion decays of the $B$ meson. In particular the study of the  
$B \to \rho \pi$ channels is used for the determination 
of the unitarity angles $\alpha$ and $\gamma$ \cite{cp}. 
These analyses make the assumption that, using cuts in the 
three invariant masses for the pion pairs, one can extract the $\rho$ 
contribution without significant background contaminations. The 
$\rho$ has spin $1$, the $\pi$ spin $0$ as well as the initial $B$, and 
therefore the $\rho$ has angular distribution $\cos ^2 \theta$ 
($\theta$ is the angle of one of the $\rho$ decay products with the 
other $\pi$ in the $\rho$ rest frame). This means that the Dalitz plot 
is mainly populated at the border, especially the corners, by this 
decay. Analyses following these lines were 
performed by the BABAR working groups \cite{babar}; MonteCarlo 
simulations, including the background from the narrow $f_0$ resonance, show 
that, with cuts at $m_{\pi\pi}=m_\rho\pm 200$ MeV, no significant contributions 
from other sources are obtained. The role of excited resonances such as 
the $\rho^\prime$ and the non--resonant background was discussed 
\cite{charles}. Finally the role of the off mass--shell $B^*$ contribution 
was discussed in \cite{bstar1,bstar2,bstar3}. 
 
The formulas derived in the present paper 
allow to estimate the contribution of the $\sigma$ which 
was not taken into account in the previous analyses. As the resonance is 
broad, part of the events from $B \to \sigma \pi \to 3 \pi$ survive 
the cuts on the invariant mass of two $\pi$ that reconstruct the mass 
of the $\rho$ (within $\pm 200$ MeV) in $B \to \rho \pi \to 3 \pi$. 
We define the integrating region in the Dalitz plot around the 
$\rho$ resonance: 
\begin{eqnarray} 
\Gamma_{eff}({\bar B}^0\to\rho^-\pi^+)&=& 
\Gamma({\bar B}^0\to\pi^+\pi^-\pi^0)\Big|_{ m_\rho-\delta 
\leq {\sqrt s}\leq m_\rho+\delta }\nonumber \\ 
\Gamma_{eff}({\bar B}^0\to\rho^+\pi^-)&=& 
\Gamma({\bar B}^0\to\pi^+\pi^-\pi^0)\Big|_{ m_\rho-\delta\leq 
{\sqrt t}\leq m_\rho+\delta }~. \nonumber 
\end{eqnarray} 
The Mandelstam variables are $s=(p_{\pi+}+p_{\pi0})^2$, 
$t=(p_{\pi-}+p_{\pi0})^2$ and we use $\delta=200$ MeV. 
 
Moreover the $\sigma$ contributes to the decay $B^\pm \to 
\pi^+ \pi^- \pi^\pm$ and only in a negligible way to the 
decay ${\bar B}^0 \to \pi^+ \pi^- \pi^0$ if its 
contribution is restricted to the experimental 
cuts that reconstruct the process ${\bar B}^0 \to \rho^\pm \pi^\mp$. 
Note that in the latter case the $\rho$ meson is charged, while 
the $\sigma$ is neutral. This means that on average the two charged 
pions will have a high invariant mass from the $\sigma$ resonance in 
the process, while in order to reconstruct a charged $\rho$ meson, a 
neutral and a charged pion have to be used. We have checked numerically 
that this is indeed correct and the $\sigma$ contribution to 
${\bar B}^0 \to \rho^\pm \pi^\mp$ (we reconstruct the $\rho$ mass 
within $\pm 200$ MeV) is three orders of magnitude smaller than 
the $\rho$ contribution (for which we find ${\mathcal B}( {\bar 
B}^0~\to~\rho^\mp\pi^\pm)~=~19.9 \times 
10^{-6}$). The $\sigma$ contribution to the charged $B$ decay is on 
the contrary a fraction of the $\rho$ one and comparable or larger 
than the contribution from the $B^*$ (see Table II). This provides a clear 
mechanism to enhance the denominator of the ratio $R$, giving a 
result closer to the experimental one: 
\begin{equation} 
R~=~\frac{ {\mathcal B}( {\bar B}^0~\to~\rho^\mp\pi^\pm) }{ {\mathcal B}( 
B^-~\to~\rho^0\pi^-)}~=~3.6 \; , 
\end{equation} 
including only the 
$\rho$ and $\sigma$ contributions, or $R\simeq 4$ including also the 
$B^*$, which however is less precisely known. A possibility to disentangle 
these contributions from the $\rho$ is to vary the experimental cuts and 
see the effect on the effective branching ratios. We find that the 
process $B^- \to \sigma \pi^-$ has a total branching ratio 
of $4.3 \times 10^{-6}$, therefore comparable to the one of the $\rho$. 
Allowing the cuts around the $\rho$ mass to be $\pm 300$ MeV, 
the $\sigma$ contribution grows to $2.7 \times 10^{-6}$, to be compared 
to what indicated in Table II for a cut of $\pm200$ MeV. 
 
\acknowledgements 
 
ADP acknowledges support from EU-TMR programme, contract 
CT98-0169. We wish to thank N.~A.~T\"ornqvist, R.~Gatto 
and G.~Nardulli for informative discussions.

\narrowtext 
\begin{table} [h] 
\caption{Estimates for the ratio $R$ beyond the factorization approximation 
(the so--called charming penguins) using different sets of input data: 
QCD sum--rules (QCDSR), lattice--QCD, quark models (QM)} 
\vspace{0.4cm} 
\begin{center} 
\begin{tabular}{cc} 
QCDSR & $R=6.3$ \\ 
lattice & $R=5.5$ \\ 
QM & $R=6.4$ \\ 
\end{tabular} 
\end{center} 
\end{table} 
 
\begin{table} 
\caption{Effective branching ratios (in units of $10^{-6}$) 
for the charged $B^-\to\pi^+\pi^-\pi^-~$ 
decay into three pions. Cuts as indicated in the text. Set I refers to the 
choice of hadronic parameter $g=0.4$ for the coupling 
$B B^* \pi$ relevant for the evaluation of the $B^*$ contribution, 
while Set II is calculated using $g=0.6$. The $\sigma$ contribution 
is less dependent on hadronic uncertainties as the coupling 
$g_{\sigma \pi^+\pi^-}$ is obtained from data.} 
\begin{tabular}{cccccc} 
$B^-\to\pi^+\pi^-\pi^-$ & $\rho$ & $\sigma$ & $B^* $ & $\rho +\sigma$& all\\ 
Set I & 3.8 & 1.5 & 0.8 & 5.5 & 4.9 \\ 
Set II & 3.8 & 1.5 & 1.8 & 5.5 & 5.1 \\ 
\end{tabular} 
\end{table} 
 
\begin{figure}[t!] 
\begin{center} 
\epsfig{bbllx=0.5cm,bblly=15cm,bburx=20cm,bbury=27cm,height=5truecm, 
        figure=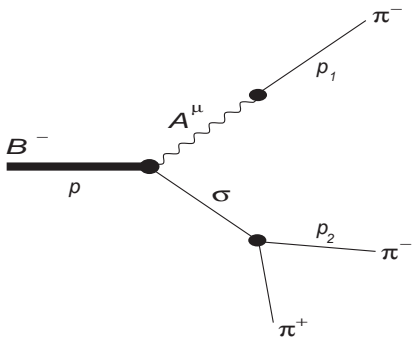} 
\caption{\label{fig:fig1} 
\footnotesize 
         Tree level diagram for a $B^-$ decaying to three pions via 
         the $\sigma$ resonance. The two identical $\pi^-$ in the 
         final state require to sum coherently two such 
         diagrams  having the momentum labels $p_1$ and $p_2$ inverted. } 
\end{center} 
\end{figure} 
\begin{figure}[t!] 
\begin{center} 
\epsfig{bbllx=0.5cm,bblly=15cm,bburx=20cm,bbury=27cm,height=5truecm, 
        figure=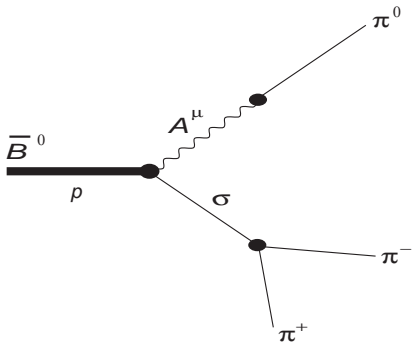} 
\caption{\label{fig:fig2} 
\footnotesize 
         As in Fig. 1 but for the $\bar{B}^0$, with no identical particles in the 
         final state.} 
\end{center} 
\end{figure} 


\begin{thebibliography}{99} 
 
\bibitem{cleo} 
Y. Kubota et al., Nucl. Inst. and Meth. A320, 66 (1992). 
 
\bibitem{babar} 
P.~F.~Harrison and H.~R.~Quinn [BABAR Collaboration], 
SLAC-R-0504 
{\it Papers from Workshop on Physics at an Asymmetric B Factory 
(BaBar Collaboration Meeting)}, 
http://www.slac.stanford.edu/pubs/slacreports/slac-r-504.html. 
 
\bibitem{kek} 
BELLE Technical Design Report, KEK-Report 95-1, April 
1995, http://bsunsrv1.kek.jp/bdocs/tdr.html; 
F.~Takasaki, 19th International Symposium on Lepton and 
Photon Interactions at High Energies (LP99), Stanford, California, 9-14 
Aug. 1999 hep-ex/9912004. 
 
\bibitem{cp} 
see for example H.~J.~Lipkin, Y.~Nir, H.~R.~Quinn and A.~Snyder, 
Phys.\ Rev.\  {\bf D44}, 1454 (1991); 
A.~E.~Snyder and H.~R.~Quinn, 
Phys.\ Rev.\  {\bf D48}, 2139 (1993); 
G.~Eilam, M.~Gronau and R.~R.~Mendel, 
Phys.\ Rev.\ Lett.\  {\bf 74}, 4984 (1995) [hep-ph/9502293]; 
I.~Bediaga, R.~E.~Blanco, C.~Gobel and R.~Mendez-Galain, 
Phys.\ Rev.\ Lett.\  {\bf 81}, 4067 (1998) 
[hep-ph/9804222]; 
R.~E.~Blanco, C.~Gobel and R.~Mendez-Galain, 
hep-ph/0007105. 
 
\bibitem{cleob} 
C.~P.~Jessop {\it et al.}  [CLEO Collaboration], 
hep-ex/0006008. 
 
\bibitem{babarb} 
B.~Aubert {\it et al.}  [BABAR Collaboration], SLAC-PUB-8537. 
 
\bibitem{gao} 
Y.~Gao and F.~Wurthwein  [CLEO Collaboration], 
hep-ex/9904008. 
 
\bibitem{BSW} 
M.~Bauer, B.~Stech and M.~Wirbel, 
Z.\ Phys.\  {\bf C34}, 103 (1987). 
 
\bibitem{bstar1} 
A.~Deandrea, R.~Gatto, M.~Ladisa, G.~Nardulli and P.~Santorelli, 
Phys.\ Rev.\  {\bf D62}, 036001 (2000) [hep-ph/0002038]; 
 
\bibitem{martinelli} 
M.~Ciuchini, E.~Franco, G.~Martinelli and L.~Silvestrini, 
Nucl.\ Phys.\  {\bf B501}, 271 (1997) 
[hep-ph/9703353]; 
M.~Ciuchini, R.~Contino, E.~Franco, G.~Martinelli and L.~Silvestrini, 
Nucl.\ Phys.\  {\bf B512}, 3 (1998); Erratum-ibid. {\bf B531}, 656 (1998) 
[hep-ph/9708222]. 
 
\bibitem{bstar2} 
A.~Deandrea, 
proceedings of 35th Rencontres de Moriond: Electroweak Interactions 
and Unified Theories, Les Arcs, France, 11-18 Mar 2000, 
hep-ph/0005014; 
 
\bibitem{bstar3} 
A.~Deandrea, R.~Gatto, M.~Ladisa, G.~Nardulli and P.~Santorelli, 
hep-ph/0007059. 
 
\bibitem{desh} 
N.~G.~Deshpande, G.~Eilam, X.~He and J.~Trampetic, 
Phys.\ Rev.\  {\bf D52}, 5354 (1995) [hep-ph/9503273]; 
 
\bibitem{pdg} 
D.~E.~Groom {\it et al.}, 
Eur.\ Phys.\ J.\  {\bf C15}, 1 (2000); 
N.~A.~Tornqvist and M.~Roos, 
Phys.\ Rev.\ Lett.\  {\bf 76}, 1575 (1996) 
[hep-ph/9511210]. 
 
\bibitem{E791} 
E.~M.~Aitala {\it et al.}  [E791 Collaboration], 
hep-ex/0007028; 
J.~Slaughter, talk at XXXVth Rencontres de Moriond on Hadronic 
interactions and QCD, http://moriond.in2p3.fr/QCD00/transparencies/ 
 
\bibitem{analisi} 
A.~D.~Polosa, N.~A.~Tornqvist, M.~D.~Scadron and V.~Elias, 
hep-ph/0005106; 
C.~Dib and R.~Rosenfeld, 
hep-ph/0006145. 
 
\bibitem{pred} 
R.~Gatto, G.~Nardulli, A.~D.~Polosa and N.~A.~Tornqvist, 
hep-ph/0007207. 
 
\bibitem{cqm} 
A.~Deandrea, N.~Di Bartolomeo, R.~Gatto, G.~Nardulli and A.~D.~Polosa, 
Phys.\ Rev.\  {\bf D58}, 034004 (1998) 
[hep-ph/9802308]; 
{\it ibid.} {\bf D59}, 074012 (1999) 
[hep-ph/9811259]; 
{\it ibid.} {\bf D61}, 017502 (2000) 
[hep-ph/9907225]; 
A.~Deandrea, R.~Gatto, G.~Nardulli and A.~D.~Polosa, 
JHEP {\bf 9902}, 021 (1999) 
[hep-ph/9901266]; 
A.~D.~Polosa, 
hep-ph/0004183. 
 
\bibitem{sanda} 
T.~Morozumi, C.~S.~Lim and A.~I.~Sanda, 
Phys.\ Rev.\ Lett.\  {\bf 65}, 404 (1990); 
Y.~Keum, U.~Nierste and A.~I.~Sanda, 
Phys.\ Lett.\  {\bf B457}, 157 (1999) 
[hep-ph/9903230]; 
K.~Terasaki, 
hep-ph/0008225. 
 
\bibitem{charles} 
J. Charles, PhD Thesis (in French); 
S. Versille, PhD Thesis (in French). 
 
\end{thebibliography}
\end{document}